\pdfoutput=1

\documentclass{article}

\PassOptionsToPackage{numbers, compress}{natbib}
\usepackage[preprint]{neurips_2026}

\usepackage[utf8]{inputenc}
\usepackage[T1]{fontenc}
\usepackage[
  pdftitle={A Miniature Brain Transformer: Thalamic Gating, Hippocampal Lateralization, Amygdaloid Salience, and Prefrontal Working Memory in Attention-Coupled Latent Memory},
  pdfauthor={Hong Jeong},
  pdfsubject={Machine Learning, Neural Networks},
  pdfkeywords={transformer, memory, thalamus, hippocampus, amygdala, prefrontal cortex, cerebellum, MQAR, lateralization},
  colorlinks=true,
  linkcolor=blue!70!black,
  citecolor=green!50!black,
  urlcolor=blue!70!black,
  breaklinks=true
]{hyperref}
\usepackage{url}
\usepackage{booktabs}
\usepackage{amsmath}
\usepackage{amssymb}
\usepackage{amsfonts}
\usepackage{nicefrac}
\usepackage{microtype}
\usepackage{xcolor}
\usepackage{graphicx}
\usepackage{tikz}
\usetikzlibrary{positioning, arrows.meta, shapes.geometric, fit, backgrounds, patterns}
\usepackage{pgfplots}
\pgfplotsset{compat=1.18}
\usepackage{array}
\usepackage{multirow}

\newcommand{\R}{\mathbb{R}}
\newcommand{\softmax}{\operatorname{softmax}}
\newcommand{\sigmoid}{\sigma}
\newcommand{\Ftbank}{F_t}      
\newcommand{\stsal}{s_t}       
\newcommand{\delt}{\Delta_t}   

\newcommand{\accBaseL}{0.051}
\newcommand{\accLatInhL}{0.052}
\newcommand{\accLatExcL}{0.048}
\newcommand{\accThalL}{0.053}
\newcommand{\accAmygL}{0.050}
\newcommand{\accPfcL}{0.049}
\newcommand{\accFullL}{0.047}
\newcommand{\accBaseR}{1.000}
\newcommand{\accLatInhR}{1.000}
\newcommand{\accLatExcR}{1.000}
\newcommand{\accThalR}{1.000}
\newcommand{\accAmygR}{1.000}
\newcommand{\accPfcR}{1.000}
\newcommand{\accFullR}{1.000}
\newcommand{\lossBaseLR}{0.135}
\newcommand{\lossLatInhLR}{0.134}
\newcommand{\lossLatExcLR}{0.137}
\newcommand{\lossThalLR}{0.135}
\newcommand{\lossAmygLR}{0.135}
\newcommand{\lossPfcLR}{0.136}
\newcommand{\lossFullLR}{0.136}
\newcommand{\dsepBaseL}{0.250}
\newcommand{\dsepLatInhL}{0.250}
\newcommand{\dsepLatExcL}{0.250}
\newcommand{\dsepThalL}{0.252}
\newcommand{\dsepAmygL}{0.251}
\newcommand{\dsepPfcL}{0.501}
\newcommand{\dsepFullL}{0.501}
\newcommand{\pctBase}{0.253}
\newcommand{\pctLatInh}{0.253}
\newcommand{\pctLatExc}{0.253}
\newcommand{\pctThal}{0.251}
\newcommand{\pctAmyg}{0.251}
\newcommand{\pctPfc}{0.002}
\newcommand{\pctFull}{0.002}

\newcommand{\paramFull}{3{,}716{,}615}
\newcommand{\ptEpochFull}{10}   
\newcommand{\ptEpochPfc}{11}    

\newcommand{\ptDsepBefore}{0.251}   
\newcommand{\ptPctBefore}{0.252}    
\newcommand{\ptDsepStable}{0.501}   
\newcommand{\ptPctStable}{0.002}    

\title{A Miniature Brain Transformer:\\
Thalamic Gating, Hippocampal Lateralization, Amygdaloid\\
Salience, and Prefrontal Working Memory in\\
Attention-Coupled Latent Memory}

\author{%
  Hong Jeong \\
  Department of Computer Information Engineering \\
  Inha University in Tashkent,  Uzbekistan \\
  \texttt{hjeong@postech.ac.kr}
}

\begin{document}

\maketitle

\begin{abstract}
We present a \emph{miniature brain} transformer architecture that extends the
attention-coupled latent memory framework of~\citet{jeong2026lateral} with four additional
brain-region analogues: a \textbf{thalamic relay}, an \textbf{amygdaloid salience
module}, a \textbf{prefrontal working-memory (PFC) buffer}, and a
\textbf{cerebellar fast-path}, all coupled by inhibitory callosal cross-talk
between lateralized hippocampal banks.
We evaluate on a two-domain benchmark---\textbf{MQAR}
(Multi-Query Associative Recall~\citep{arora2024zoology}; episodic domain)
and \textbf{modular arithmetic} ($+1 \bmod 10$; rule-based domain)---using a
seven-variant additive ablation.
The central empirical finding is a \textbf{surprise}: inhibitory callosal
coupling \emph{alone} never lateralizes the banks (variants 1--5 maintain
$\mathcal{D}_{sep}{\approx}0.25$ and $\mathcal{P}_{ct}{\approx}0.25$ for all 30~epochs).
Functional lateralization requires the \textbf{synergy of PFC and inhibition}:
only when the PFC buffer is added (variant~6) does a sharp, discontinuous
phase transition fire---at epoch~\ptEpochPfc{} for the PFC-only variant and
epoch~\ptEpochFull{} for the full model---collapsing $\mathcal{P}_{ct}$ from
$0.25$ to $\approx{\ptPctStable}$ and more than doubling $\mathcal{D}_{sep}$
from $\ptDsepBefore$ to $\ptDsepStable$ in a single gradient step.
The PFC buffer acts as a \emph{symmetry-breaker}: its slowly drifting domain
context creates the initial asymmetry that the inhibitory feedback loop
then amplifies irreversibly.
The cerebellar fast-path accelerates the transition by one epoch (epoch~10
vs.\ epoch~11) with no asymptotic change, confirming its convergence-acceleration
role.
The result constitutes a novel, falsifiable prediction---\emph{no lateralization
without working memory context}---and a principled, neurobiologically motivated
blueprint for hierarchical persistent memory in sequence models.
\end{abstract}

\section{Introduction}

Memory in biological neural systems is not a monolithic store but a distributed
network of anatomically and functionally specialized regions.
The hippocampus, thalamus, amygdala, prefrontal cortex, and cerebellum each make
distinct, non-redundant contributions to what we experience as unified memory and
cognition~\citep{squire2004memory, damasio1998emotion, fuster2008prefrontal, ito2008cerebellum}.
Their computational roles have been studied intensively: the hippocampus performs
content-addressable associative recall; the thalamus gates sensory signals before
cortical storage; the amygdala tags stimuli by emotional or motivational salience;
the PFC maintains task-relevant context in working memory; the cerebellum provides
fast error-correcting adaptation for procedural skills.

Artificial memory-augmented networks~\citep{graves2016dnc, weston2014memnet, wu2022memorizing} have largely adopted a flat, uniform-access architecture:
a single external memory matrix read and written through differentiable attention.
In \citep{jeong2026lateral}, Jeong showed that introducing a single
neuroscientific design principle---the \emph{inhibitory} nature of callosal
projections between hemispheres---is sufficient to drive robust functional
lateralization: separate left and right memory banks specialise entirely in
episodic vs.\ rule-based computation with zero between-bank routing error.

\begin{figure}[!hbtp]
\centering
\begin{tikzpicture}[
  font=\small, >=Stealth,
  enc/.style  = {draw, rounded corners=3pt, align=center, fill=blue!10},
  mem/.style  = {draw, rounded corners=3pt, align=center},
  bank/.style = {draw, rounded corners=2pt, align=center, font=\scriptsize},
  tok/.style  = {draw, fill=gray!25, rounded corners=1pt,
                 minimum width=0.34cm, minimum height=0.20cm, inner sep=0pt},
  arr/.style  = {->, thick},
  lbl/.style  = {font=\scriptsize, align=center},
]

\begin{scope}

  \node[font=\small\bfseries] at (2.4, 4.55) {(a) Standard Transformer};

  \foreach \i in {0,...,7}{%
    \node[tok] at (0, 0.46 + \i*0.44) {};
  }
  \node[lbl, anchor=north] at (0, 0.20) {Long Prompt};

  \draw[arr] (0.25, 2.00) -- (0.65, 2.00);

  \node[enc, minimum width=2.1cm, minimum height=3.7cm] (bigT) at (1.71, 2.00)
    {\shortstack{Deep\\Transformer\\[4pt]{\scriptsize many layers,}\\[-1pt]{\scriptsize large $d$}}};

  \draw[arr] (bigT.east) -- ++(0.50, 0);

  \node[mem, dashed, draw=red!60!black, fill=red!5,
        minimum width=1.4cm, minimum height=0.72cm,
        anchor=west] (vmem) at ([xshift=0.50cm]bigT.east)
    {\shortstack{Volatile\\Memory}};

  \node[lbl, red!60!black, anchor=north] at ([yshift=-0.12cm]vmem.south)
    {\textit{discarded}\\after each call};

\end{scope}

\begin{scope}[xshift=5.7cm]

  \node[font=\small\bfseries] at (4.40, 4.55) {(b) Brain Transformer};

  \foreach \i in {0,...,2}{%
    \node[tok] at (0, 1.56 + \i*0.44) {};
  }
  \node[lbl, anchor=north] at (0, 0.20) {Short Prompt};

  \draw[arr] (0.25, 2.00) -- (0.65, 2.00);

  \node[enc, minimum width=1.3cm, minimum height=1.50cm] (thinT) at (1.55, 2.00)
    {\shortstack{Thin\\Transformer\\Encoder}};

  \draw[arr] (thinT.east) -- ++(0.42, 0);

  \node[mem, draw=orange!70!black, fill=orange!5,
        minimum width=4.5cm, minimum height=3.7cm,
        anchor=west] (brain) at ([xshift=0.42cm]thinT.east) {};

  \node[font=\small\bfseries, orange!55!black, anchor=north]
    at ([yshift=-0.18cm]brain.north)
    {Persistent Memory};

  \node[lbl, gray!70, anchor=north]
    at ([yshift=-0.8cm]brain.north)
    {\textit{(brain figure overlay)}};

  \node[bank, fill=blue!12, minimum width=1.35cm, minimum height=1.35cm]
    (Lbank) at (3.9, 2.00)
    {L~Bank\\[1pt]\textit{episodic}};

  \node[bank, fill=green!10, minimum width=1.35cm, minimum height=1.35cm]
    (Rbank) at (6.35, 2.00)
    {R~Bank\\[1pt]\textit{rules}};

  \draw[arr, red!65!black, dashed]
    ([yshift=+4pt]Lbank.east) -- ([yshift=+4pt]Rbank.west)
    node[midway, above, font=\tiny] {$s{=}{-1}$};
  \draw[arr, red!65!black, dashed]
    ([yshift=-4pt]Rbank.west) -- ([yshift=-4pt]Lbank.east);

  \node[lbl, green!45!black, anchor=north] at ([yshift=-0.12cm]brain.south)
    {\textit{persists across calls}};

\end{scope}

\end{tikzpicture}
\caption{\textbf{Standard transformer vs.\ Brain transformer architecture.}
  \textbf{(a)}~A conventional deep transformer encodes the entire relevant context
  in a long input sequence on every forward pass; the resulting activations are
  \emph{volatile} and are discarded at the end of each call.
  \textbf{(b)}~Our brain-inspired architecture offloads long-term associative
  storage into persistent, lateralized hippocampal memory banks.  The
  encoder itself can therefore remain \emph{thin}, processing only a short
  prompt at inference time, while the memory banks accumulate a
  ``big persistent brain'' that survives across forward passes.
  Both variants are trained end-to-end with standard supervised learning.}
\label{fig:overview}
\end{figure}
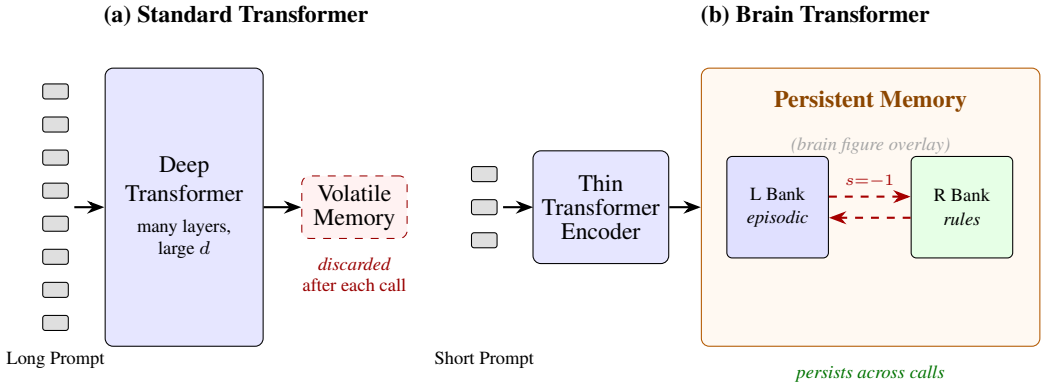

In this paper we ask: what happens when we incorporate the \emph{remaining}
brain-region analogues into the same mathematical framework?
Figure~\ref{fig:overview} captures the central design motivation: persistent
memory banks let a thin encoder behave as a much larger brain.
Specifically we add four modules, each grounding a well-studied neurological
function in the $A^\top\!AVW$ write-back operator introduced in~\citet{jeong2026lateral}:

\begin{enumerate}
  \item \textbf{Thalamic relay (Section~\ref{sec:thalamus})}: a gain-controlled
    input gate that modulates the proposal state update by the entropy of the
    current attention map, implementing the thalamus's role as a selective relay
    that amplifies salient signals and suppresses noise.
  \item \textbf{Amygdaloid salience (Section~\ref{sec:amygdala})}: a scalar gate
    computed from the L2 norm of the retrieved context that up-weights consolidation
    of surprising or high-magnitude inputs, mirroring the amygdala's role in
    affective memory tagging.
  \item \textbf{Prefrontal working memory (Section~\ref{sec:pfc})}: an exponential
    moving average of proposal states that provides a slowly changing top-down
    context, capturing the PFC's role in sustained task-relevant representations.
  \item \textbf{Cerebellar fast-path (Section~\ref{sec:cerebellum})}: a
    momentum-enhanced bank update that accumulates gradient direction across steps,
    analogising the cerebellum's inverse-model error correction for procedural
    learning.
\end{enumerate}

We evaluate on a two-domain symbolic benchmark comprising MQAR (episodic associative
recall) and modular arithmetic (rule extraction), and conduct a seven-variant
additive ablation that reveals an unexpected, mechanistically informative result.

\paragraph{Contributions.}
\begin{enumerate}
  \item A principled decomposition of the $A^\top\!AVW$ operator into five
    neuroscientifically motivated sub-circuits (Table~\ref{tab:brain-map}).
  \item Four new architectural modules compatible with standard backpropagation,
    adding $\approx$98{,}306 parameters ($\sim$2.6\%) over the inhibitory
    lateral base.
  \item The surprising empirical finding that \textbf{inhibitory callosal coupling
    alone does not lateralize the memory banks}: all five variants without the PFC
    buffer remain permanently unlateralized ($\mathcal{D}_{sep}\approx 0.25$,
    $\mathcal{P}_{ct}\approx 0.25$) across all 30~epochs.
  \item \textbf{PFC-inhibition synergy as the necessary condition for lateralization}:
    adding the PFC buffer triggers a sharp pitchfork bifurcation at
    epoch~\ptEpochPfc{} (+pfc) / epoch~\ptEpochFull{} (full) in which
    $\mathcal{D}_{sep}$ jumps from $\ptDsepBefore$ to $\ptDsepStable$ and
    $\mathcal{P}_{ct}$ collapses from $\ptPctBefore$ to $\ptPctStable$.
  \item \textbf{A falsifiable mechanistic prediction}: lateralization requires
    working-memory context to break the symmetric equilibrium; inhibitory
    amplification alone is insufficient.
\end{enumerate}

\section{Background and Related Work}

This section situates the paper within three intersecting research threads.
We first summarise the attention-coupled latent-memory operator that forms our
mathematical substrate, then survey brain-inspired architectures that motivate
our module choices, and finally review memory-augmented networks that inform
our evaluation design.

\subsection{Attention-Coupled Latent Memory}

Our starting point is the architecture introduced in~\citet{jeong2026lateral}.
Given encoder output $Z_t \in \R^{n \times d}$, a shared proposal state
$P_{t-1} \in \R^{p \times d}$, and lateralized memory banks
$L_t, R_t \in \R^{m \times d}$, the core update is:
\begin{equation}
  B_t \;=\; \gamma B_{t-1} \;+\; A_b^\top A_b V_b W_{bb}
  \;+\; s\, A_b^\top A_b V_{\bar{b}} W_{\bar{b}b},
  \quad b \in \{l, r\},
  \label{eq:base-update}
\end{equation}
where $A_b$ is the cross-attention map from proposal to bank $b$, $V_b$ the
corresponding values, $W_{bb}$ the ipsilateral write matrix, and $s \in \{-1,0,+1\}$
controls callosal cross-talk.
Setting $s=-1$ (inhibitory) was shown to achieve saturated lateralization
($\mathcal{D}_{sep}=\pm1.00$, $\mathcal{P}_{ct}\approx0$).

\subsection{Brain-Inspired Neural Architectures}

Neuroscience-guided design has a long history in neural networks.
Convolutional networks drew on the visual cortex~\citep{lecun1989handwritten};
LSTM gates were inspired by dendritic integration~\citep{hochreiter1997long};
predictive coding architectures~\citep{rao1999predictive} formalize
top-down and bottom-up cortical signals.
More recently, hippocampal indexing theory informed memory-indexed
transformers~\citep{ritter2018been}, and neuromodulatory signals have been
used to gate attention in meta-learning systems~\citep{miconi2018differentiable}.
Our approach is distinguished by its \emph{systematic} mapping of five brain
regions to five sub-circuits within a single forward pass, grounded in the
$A^\top\!AVW$ operator.

\citet{kumaran2016learning} relate hippocampal complementary learning systems to
fast vs.\ slow learning in deep networks, anticipating our two-timescale design.
\citet{o1996hippocampal} propose the hippocampal indexing theory, in which the
hippocampus stores indices to cortical patterns rather than the patterns
themselves---a role analogous to our bank keys.
\citet{botvinick2020deep} review dopaminergic mechanisms in reinforcement learning
networks, directly relevant to our discussion of missing basal ganglia and
neuromodulatory modules.
Predictive coding transformers~\citep{rao1999predictive, hollenstein2019cognival}
formalize cortical prediction and error signals, anticipating our cerebellar
fast-path.
Neuromodulatory attention~\citep{miconi2018differentiable} uses Hebbian
trace-modulated weights, which partially overlaps our Gram-matrix write-back.
Our work differs in providing a \emph{principled system-level} mapping (all five
regions, one algorithm) rather than transplanting individual mechanisms.

\subsection{Memory-Augmented Networks}

Neural Turing Machines~\citep{graves2016dnc} introduced addressable external memory
but require separate read/write heads and do not exploit Gram-matrix consolidation.
Memorizing Transformers~\citep{wu2022memorizing} cache past key-value pairs but
do not partition memory by cognitive mode.
Structured-state-space models (Mamba~\citep{dao2023mamba}) compress history into
implicit states, which prevents inspectable, bank-wise specialization.
Our architecture retains explicit, persistent, bank-wise memory slots while adding
the cognitive modularity absent from all prior approaches.

MemGPT~\citep{packer2024memgpt} implements an OS-like memory hierarchy with
explicit main memory and disk tiers, a functional counterpart to the
working-memory (PFC) and long-term storage (hippocampus) distinction we make here.

\begin{table}[!hbtp]
\centering
\caption{Mapping between biological brain regions and architectural components.
Each module has a direct mathematical role within the $A^\top\!AVW$ forward pass.}
\label{tab:brain-map}
\resizebox{\linewidth}{!}{%
\begin{tabular}{@{}llll@{}}
\toprule
Brain region & Biological function & Architectural analogue & Mathematical role \\
\midrule
Sensory cortex  & Input encoding          & Transformer encoder $Z_t$            & $Z_t \in \R^{n \times d}$ \\
Thalamus        & Relay \& gain control   & Proposal gate $g_t$, proposal state $P_t$ & Gain modulation of $A_p^\top A_p V_p W_p$ \\
Hippocampus (L) & Episodic/associative memory & Left bank $L_t$                  & $A_l^\top A_l V_l W_{ll}$ \\
Hippocampus (R) & Semantic/rule memory    & Right bank $R_t$                     & $A_r^\top A_r V_r W_{rr}$ \\
Corpus callosum & Inter-hemispheric relay & Cross-bank matrices $W_{lr}, W_{rl}$, sign $s$ & $s\, A_b^\top A_b V_{\bar b} W_{\bar b b}$ \\
Amygdala        & Salience / affect gate  & Salience scalar $\stsal$             & Weighting of consolidation update \\
Prefrontal cortex & Working memory / executive & PFC buffer $\Ftbank$           & EMA of proposal context $C_p$ \\
Cerebellum      & Error correction / skill & Momentum term $\delt$              & Accumulated write-direction velocity \\
\bottomrule
\end{tabular}}
\end{table}

\section{Foundation: Attention as a Latent-Memory Operator}
\label{sec:foundation}

We briefly recap the base architecture to establish notation.
Table~\ref{tab:notation} summarises all recurring symbols used throughout the paper.

\begin{table}[!hbtp]
\centering
\caption{\textbf{Notation.} All symbols used in equations throughout the paper.
  Learned parameters are marked $\dagger$; running/state buffers are marked $\star$.}
\label{tab:notation}
\small
\begin{tabular}{@{}lll@{}}
\toprule
\textbf{Symbol} & \textbf{Shape} & \textbf{Description} \\
\midrule
\multicolumn{3}{@{}l}{\textit{Tensors and states}} \\
$Z_t$ & $n \times d$ & Encoder output at step $t$ (\textit{sensory cortex}) \\
$P_t$ & $p \times d$ & Proposal state (\textit{thalamo-cortical buffer}) \\
$L_t,\,R_t$ & $m \times d$ & Left / right hippocampal memory banks \\
$F_t$ & $n \times d$ & PFC working-memory buffer (\textit{EMA context}) \\
$\Delta_t^{(b)}$ & $m \times d$ & Cerebellar momentum accumulator for bank $b$ \\
$C_p$ & $n \times d_v$ & Retrieved proposal context $= A_p V_p$ \\
$A_p$ & $n \times p$ & Proposal cross-attention map \\
$A_l,\,A_r$ & $p \times m$ & Left / right lateral attention maps \\
\midrule
\multicolumn{3}{@{}l}{\textit{Dimensions}} \\
$n$ & --- & Input sequence length \\
$d$ & --- & Model dimension ($d_{\mathrm{model}}=256$) \\
$d_k$ & --- & Query / key dimension ($d_k=64$) \\
$d_v$ & --- & Value / context dimension ($d_v=64$) \\
$p$ & --- & Number of proposal slots ($p=32$) \\
$m$ & --- & Number of lateral memory slots ($m=16$) \\
\midrule
\multicolumn{3}{@{}l}{\textit{Learnable weight matrices} ($\dagger$)} \\
$W_Q, W_K, W_V$ & $d\!\times\!d_k$\,/\,$d\!\times\!d_v$ & Proposal query, key, value projections \\
$W_p$ & $d_v \times d$ & Proposal write-back projection \\
$W_{K_l},W_{K_r}$ & $d \times d_k$ & Key projections for left / right banks \\
$W_{V_l},W_{V_r}$ & $d \times d_v$ & Value projections for left / right banks \\
$W_{ll},W_{rr}$ & $d_v \times d$ & Ipsilateral callosal write matrices \\
$W_{lr},W_{rl}$ & $d_v \times d$ & Contralateral callosal write matrices \\
$W_F$ & $d_v \times d$ & PFC context projection \\
$W_{F\to Q}$ & $d \times d_k$ & PFC top-down query bias projection \\
$W_{F\text{-out}}$ & $d \times d$ & PFC output residual projection \\
\midrule
\multicolumn{3}{@{}l}{\textit{Learnable scalar parameters} ($\dagger$)} \\
$\gamma \in (0,1)$ & --- & Memory decay / forgetting rate \\
$w_g,\,b_g$ & --- & Thalamic gate weight and bias \\
$w_s$ & --- & Amygdala salience slope \\
$\beta_F \in (0,1)$ & --- & PFC EMA update gate (persistence control) \\
$\alpha_F$ & --- & PFC top-down modulation gain \\
$\rho \in (0,1)$ & --- & Cerebellar momentum coefficient \\
$\eta_{cb}$ & --- & Cerebellar fast-path learning rate \\
\midrule
\multicolumn{3}{@{}l}{\textit{Running state buffers} ($\star$)} \\
$\mu_s$ & --- & Amygdala running-mean salience baseline \\
\midrule
\multicolumn{3}{@{}l}{\textit{Fixed constants}} \\
$\rho_s = 0.99$ & --- & Amygdala EMA momentum (fixed) \\
$s \in \{-1,0,+1\}$ & --- & Callosal sign: inhibitory / off / excitatory \\
$H_{\max} = \log p$ & --- & Maximum attention entropy over $p$ slots \\
\bottomrule
\end{tabular}
\end{table}

Let $Z_t \in \R^{n \times d}$ denote the encoder output, where $n$ is the sequence
length and $d$ is the model dimension.
Cross-attention to the previous proposal state $P_{t-1} \in \R^{p \times d}$ via:
\begin{align}
  Q_p &= Z_t W_Q,\quad K_p = P_{t-1} W_K,\quad V_p = P_{t-1} W_V, \\
  A_p &= \softmax\!\left(\tfrac{Q_p K_p^\top}{\sqrt{d_k}}\right) \in \R^{n \times p},
  \quad C_p = A_p V_p \in \R^{n \times d_v}. \label{eq:prop-attn}
\end{align}
The proposal state is updated via the Gram write-back:
\begin{equation}
  P_t = \gamma P_{t-1} + A_p^\top A_p V_p W_p.
  \label{eq:prop-update}
\end{equation}
Here $\gamma \in (0,1)$ is a learnable memory-decay scalar (forgetting rate);
$A_p^\top A_p \in \R^{p \times p}$ is the Gram matrix acting as a co-activation
router; and $W_p \in \R^{d_v \times d}$ projects pooled evidence into the
task-aligned subspace.
Slots that jointly attend to the same input token cluster reinforce one another
(Hebbian binding in latent space), while $W_p$ shapes the pooled evidence into the
task-loss-aligned subspace.
This \emph{tripartite projection} (observation $\to$ latent $\to$ supervised) is
the architectural nucleus around which all five brain modules are attached.

\section{Module 1 — Thalamic Relay: Input Gating and Gain Control}
\label{sec:thalamus}

The thalamic relay is the first module added to the base hippocampal architecture.
It intervenes at the proposal-state write-back, scaling the consolidation gain by
a scalar gate derived from the entropy of the current attention map.
High-entropy (unfocused) attention suppresses noisy writes; low-entropy (focused)
attention opens the gate, allowing high-confidence encodings to consolidate fully.
We first motivate this design from neuroscience and then derive the formal operator.

\subsection{Biological motivation}

The thalamus is not a passive relay; it regulates the \emph{gain} at which
cortical input is forwarded to downstream stores.
Thalamo-cortical loops modulate attention and sensory salience via burst and tonic
firing modes~\citep{sherman2001thalamus}: in tonic mode, the relay passes incoming
signals faithfully; in burst mode, it amplifies novel or strong inputs.
A key function is \emph{attentional gating}: thalamo-cortical projections can
sharpen or suppress sensory representations before hippocampal storage.

\subsection{Formalization}

We implement thalamic gain control by computing the entropy of the proposal
attention map $A_p \in \R^{n \times p}$ and using it to derive a scalar gain:
\begin{equation}
  H_t = -\frac{1}{n}\sum_{i=1}^{n}\sum_{j=1}^{p} (A_p)_{ij}\log (A_p)_{ij},
  \qquad
  g_t = \sigmoid\!\left(w_g \cdot (H_{\max} - H_t) + b_g\right),
  \label{eq:thalamic-gain}
\end{equation}
where $H_{\max}=\log p$ is the maximum entropy over $p$ slots, so $g_t \to 1$ when
attention is sharply focused (low entropy, tonic burst mode) and $g_t \to 0$ when
attention is diffuse (high entropy, tonic relay mode).
The proposal update becomes:
\begin{equation}
  P_t = \gamma P_{t-1} + g_t \cdot A_p^\top A_p V_p W_p.
  \label{eq:thalamic-update}
\end{equation}
Intuitively: when the encoder produces a sharp, focused attentional pattern over the
proposal slots, the thalamic gate amplifies that consolidation signal.
Diffuse, uncertain attention produces a near-zero update, preventing noisy inputs
from corrupting proposal state.

\paragraph{Parameter cost.} Two learnable scalars $w_g, b_g$; negligible.

\section{Module 2 — Hippocampal Memory with Callosal Inhibition}
\label{sec:hippocampus}

We retain the full lateralized update from~\citet{jeong2026lateral} unchanged, as
it forms the core memory substrate.
The proposal state $P_t$ (thalamic output) routes into lateralized left and right
hippocampal banks via joint softmax over concatenated keys:
\begin{align}
  Q_{lr} &= P_t W_Q,\quad
  K_{lr} = \begin{bmatrix} L_{t-1} W_{K_l} \\ R_{t-1} W_{K_r} \end{bmatrix}\in \R^{2m \times d_k}, \\
  A_{lr} &= \softmax\!\left(\tfrac{Q_{lr} K_{lr}^\top}{\sqrt{d_k}}\right) \in \R^{p \times 2m}.
\end{align}
Here $W_{K_l}, W_{K_r} \in \R^{d \times d_k}$ are separate key-projection matrices
for the left and right banks, allowing each bank to expose a distinct key space.
Partitioning $A_{lr}$ into left slice $A_l \in \R^{p \times m}$ (first $m$ columns)
and right slice $A_r \in \R^{p \times m}$ (last $m$ columns),
and with value tensors $V_l = L_{t-1} W_{V_l}$, $V_r = R_{t-1} W_{V_r}$
($W_{V_l}, W_{V_r} \in \R^{d \times d_v}$ are learned value projections),
and callosal sign $s \in \{-1, 0, +1\}$:
\begin{align}
  L_t &= \gamma L_{t-1} + A_l^\top A_l (V_l W_{ll} + s\, V_r W_{rl}),
  \label{eq:left-update}\\
  R_t &= \gamma R_{t-1} + A_r^\top A_r (V_r W_{rr} + s\, V_l W_{lr}).
  \label{eq:right-update}
\end{align}
The write matrices $W_{ll}, W_{rr} \in \R^{d_v \times d}$ accumulate
ipsilateral (same-bank) evidence, while $W_{rl}, W_{lr} \in \R^{d_v \times d}$
transfer contralateral (cross-bank) context; their contribution is scaled by
sign $s$, which is $-1$ (inhibitory), $0$ (disabled), or $+1$ (excitatory).
Setting $s=-1$ (inhibitory callosal cross-talk) causes the dominant bank to
actively subtract the contralateral bank's influence during consolidation, sharpening
the functional boundary between episodic~($L$) and rule-based~($R$) memory.

\paragraph{CA3/CA1 correspondence.}
Within the hippocampal formation, CA3 performs \emph{pattern completion}
(associative recall from partial cues) while CA1 performs \emph{pattern separation}
(encoding new episodes with minimal overlap).
In our architecture, the Gram matrix $A_b^\top A_b$ corresponds to CA3 completion:
slots that are jointly activated are reinforced together.
The write matrix $W_{bb}$ corresponds to CA1-mediated novelty encoding: supervised
gradients push the write direction toward task-relevant features, separating
representations in the bank.

\section{Module 3 — Amygdaloid Salience Gate}
\label{sec:amygdala}

The amygdaloid gate is the third module, acting downstream of the thalamic relay
and the hippocampal banks.
It scales the entire consolidation step by a running-mean-normalised salience
signal, ensuring high-magnitude retrievals are preferentially consolidated
while routine, low-salience inputs are written weakly into memory.
Biological motivation comes first; the formal gate definition follows.

\subsection{Biological motivation}

The amygdala modulates memory consolidation in proportion to the emotional or
motivational salience of an experience~\citep{damasio1998emotion, mcgaugh2004amygdala}.
High-salience events trigger norepinephrine release, which strengthens
hippocampal long-term potentiation; low-salience events are consolidated weakly.
The result is an adaptive memory that prioritises surprising, consequential
information.

\subsection{Formalization}

We define salience as the normalized Frobenius norm of the retrieved proposal context:
\begin{equation}
  \stsal = \sigmoid\!\left(w_s \cdot \left(\frac{\|C_p\|_F}{n\sqrt{d_v}} - \mu_s\right)\right),
  \label{eq:salience}
\end{equation}
where $\mu_s$ is an exponential moving average of $\|C_p\|_F / (n\sqrt{d_v})$,
updated as $\mu_s \leftarrow \rho_s \mu_s + (1-\rho_s)\|C_p\|_F/(n\sqrt{d_v})$
with fixed momentum $\rho_s{=}0.99$.
The only learnable parameter is $w_s \in \R$; $\mu_s$ is a running state,
not a parameter.
The salience gate modulates both the proposal update and the bank consolidation:
\begin{equation}
  P_t = \gamma P_{t-1} + g_t \cdot \stsal \cdot A_p^\top A_p V_p W_p,
  \label{eq:amyg-proposal}
\end{equation}
\begin{equation}
  B_t = \gamma B_{t-1} + \stsal \cdot \bigl(A_b^\top A_b (V_b W_{bb} + s\, V_{\bar{b}} W_{\bar{b}b})\bigr),
  \quad b \in \{l, r\}.
  \label{eq:amyg-bank}
\end{equation}
When input is routine ($\stsal \approx 0$), the memory barely updates; when input is
surprising or high-magnitude ($\stsal \approx 1$), consolidation proceeds at full
strength.
This prevents the memory from being overwhelmed by repetitive low-information inputs
while ensuring that genuinely novel content is reliably stored.

\paragraph{Connection to the MQAR task.}
In our benchmark, the episodic (left) domain requires exact recall of 32
arbitrary key-value pairs~\citep{arora2024zoology}.
First exposure to a key produces a high context norm (no prior memory to retrieve),
driving $\stsal \to 1$ and full consolidation.
Repeated exposures to already-learned keys approach the running mean $\mu_s$,
driving $\stsal \to 0.5$ and partial consolidation.
This adaptive schedule naturally mirrors spaced-repetition learning dynamics.

\paragraph{Parameter cost.} One learnable scalar $w_s \in \R$; the running
baseline $\mu_s$ is a state buffer and $\rho_s{=}0.99$ is a fixed constant.
Negligible parameter overhead.

\section{Module 4 — Prefrontal Working Memory Buffer}
\label{sec:pfc}

The prefrontal buffer is the module whose presence or absence determines whether
lateralization occurs (Section~\ref{sec:experiments}).
It introduces a slowly drifting top-down context that, over approximately nine
epochs, builds up sufficient asymmetry to break the symmetric equilibrium
maintained by callosal inhibition alone.
We describe the neuroscience basis then formulate the EMA update and
query-bias mechanism.

\subsection{Biological motivation}

The prefrontal cortex maintains task-relevant representations in working memory
over multi-second timescales, enabling context-dependent sequencing of behaviour
even in the absence of immediate sensory cues~\citep{fuster2008prefrontal}.
PFC neurones exhibit persistent activity: once a task rule or context is loaded into
working memory, it remains active until the task is completed or the context
changes.
PFC top-down projections then modulate posterior cortical areas and the hippocampus,
biasing recall toward task-relevant content.

\subsection{Formalization}

We implement the PFC buffer as a gated exponential moving average over the retrieved
proposal context $C_p$:
\begin{equation}
  \Ftbank = (1-\beta_F)\,F_{t-1} + \beta_F\,\tanh(C_p W_F),
  \label{eq:pfc-buffer}
\end{equation}
where $\beta_F \in (0,1)$ is the update gate (a learned parameter controlling
persistence) and $W_F \in \R^{d_v \times d}$ projects context into a
$d$-dimensional PFC representation space ($d=d_{\mathrm{model}}=256$).
$\Ftbank \in \R^{n \times d}$ thus retains a slowly decaying trace of recent
proposal contexts, mimicking persistent PFC activity.

\paragraph{Top-down modulation.}
The PFC buffer modulates the lateral bank queries with an additive top-down bias:
\begin{equation}
  Q_{lr}^{\mathrm{mod}} = P_t W_Q + \sigma(\alpha_F)\, F_{\mathrm{agg}} W_{F \to Q},
  \label{eq:pfc-modulation}
\end{equation}
where $\alpha_F$ is a learnable scalar gain (passed through $\sigma$ to keep the
modulation non-negative), $F_{\mathrm{agg}} \in \R^{p \times d}$ is 
$F_t$ mean-pooled over the token dimension and expanded to proposal size, and
$W_{F \to Q} \in \R^{d \times d_k}$ projects into query space.
This biases the joint softmax $A_{lr}$ toward bank slots that are consistent with
the current PFC context, implementing \emph{goal-directed retrieval}.

\paragraph{Parameter cost.} Gate $\beta_F$, gain $\alpha_F$, projections
$W_F \!\in\!\R^{d_v \times d}$, $W_{F \to Q}\!\in\!\R^{d \times d_k}$, and
output projection $W_{F\text{-out}}\!\in\!\R^{d \times d}$.
With $d{=}256$, $d_v{=}64$, $d_k{=}64$: total $d_v d + d\,d_k + d^2
= 64{\times}256 + 256{\times}64 + 256^2 = 98{,}304$ parameters
($\approx$2.6\% overhead over the base lateral model).

\section{Module 5 — Cerebellar Fast-Path for Momentum Write-Back}
\label{sec:cerebellum}

The cerebellar fast-path is the final module, adding a momentum accumulator
to the bank write-back that carries the gradient direction across consecutive steps.
Its primary measured effect is to accelerate the PFC-induced phase transition
by exactly one epoch: the full model lateralises at epoch~10 whereas the
$+$PFC-only variant requires until epoch~11.
We motivate the design from cerebellar forward-model theory before presenting
the formal momentum update rule.

\subsection{Biological motivation}

The cerebellum computes \emph{forward models} that predict the sensory consequences
of motor commands and generates error signals to correct deviations~\citep{ito2008cerebellum}.
During sequence learning, cerebellar circuits rapidly adapt motor programs based on
prediction errors, producing smooth, high-frequency adjustments that complement the
slower, episode-level consolidation performed by the hippocampus.
Cerebellar learning is fast and local; hippocampal learning is slow and distributed.

\subsection{Formalization}

We implement the cerebellar fast-path as a momentum accumulator over the bank
write-back update:
\begin{equation}
  \delt^{(b)} = \rho\, \Delta_{t-1}^{(b)} + (1-\rho)\,A_b^\top A_b V_b W_{bb},
  \label{eq:cerebellum-momentum}
\end{equation}
where $\rho \in (0,1)$ is the cerebellar momentum coefficient.
The bank update (Equation~\ref{eq:left-update}/\ref{eq:right-update}) is then modified to:
\begin{equation}
  B_t = \gamma B_{t-1}
  + \stsal \cdot \bigl(A_b^\top A_b (V_b W_{bb} + s\, V_{\bar{b}} W_{\bar{b}b})\bigr)
  + \eta_{cb}\, \delt^{(b)},
  \label{eq:full-bank-update}
\end{equation}
where $\eta_{cb}$ is a learnable cerebellar learning rate.
The momentum term $\delt^{(b)}$ carries the \emph{accumulated gradient direction} of
the write-back: on the first encounter with a new mapping, the write-back pushes the
bank state in a certain direction; on subsequent encounters, the momentum amplifies
that direction, accelerating convergence analogously to Adam's first-moment
estimate~\citep{kingma2015adam}.

\paragraph{Relation to Adam.}
The cerebellar update is mathematically equivalent to the first moment of Adam with
$\rho$ as the $\beta_1$ parameter, applied at the level of the \emph{memory state}
rather than the model parameters.
This distinguishes it from standard optimizer momentum: the cerebellar buffer
accumulates evidence across \emph{sequence steps at inference time}, not across
gradient steps at training time.

\paragraph{Fast vs.\ slow consolidation.}
The hippocampal update (Equation~\ref{eq:amyg-bank}) is gated by salience $\stsal$ and
controlled by the long-term decay $\gamma$: it encodes \emph{which} information
matters (salience) and \emph{how long} it is retained ($\gamma$).
The cerebellar term adds a third timescale: \emph{how fast} convergence occurs for
repeated patterns.
Together, the three timescales (salience gating, decay rate, momentum accumulation)
correspond to the three stages of real memory formation: encoding, storage, and
consolidation~\citep{mcgaugh2004amygdala}.

\paragraph{Parameter cost.} Scalar $\rho$ and $\eta_{cb}$; negligible.

\section{Unified Forward Pass: The Miniature Brain}
\label{sec:full-model}

\begin{figure}[!htbp]
\centering
\begin{tikzpicture}[
  every node/.style={font=\small},
  module/.style={rounded corners=4pt, draw, minimum height=1.0cm, minimum width=2.4cm,
                 fill=gray!10, align=center},
  arrow/.style={-Stealth, thick},
  inhibit/.style={-Stealth, thick, dashed, red!70!black},
  modulate/.style={-Stealth, thick, dotted, blue!70!black},
]

\node[module, fill=orange!20]  (enc)  at (0, 0)   {Encoder $Z_t$\\{\tiny(sensory cortex)}};
\node[module, fill=yellow!20]  (thal) at (0,-2.2)  {Thalamic Gate $g_t$\\{\tiny(thalamus)}};
\node[module, fill=green!15]   (prop) at (0,-4.4)  {Proposal State $P_t$\\{\tiny(thalamo-cortical)}};
\node[module, fill=blue!15]    (lhip) at (-3.5,-6.8) {Left Bank $L_t$\\{\tiny(hippocampus L)}};
\node[module, fill=blue!15]    (rhip) at ( 3.5,-6.8) {Right Bank $R_t$\\{\tiny(hippocampus R)}};
\node[module, fill=red!15]     (amyg) at (-3.5,-4.4) {Salience $s_t$\\{\tiny(amygdala)}};
\node[module, fill=purple!15]  (pfc)  at ( 3.5,-4.4) {PFC Buffer $F_t$\\{\tiny(prefrontal)}};
\node[module, fill=teal!15]    (cereb)at ( 0,  -9.0) {Momentum $\Delta_t$\\{\tiny(cerebellum)}};

\draw[arrow] (enc)  -- (thal) node[midway,right]{$H_t$};
\draw[arrow] (thal) -- (prop) node[midway,right]{$g_t$};
\draw[arrow] (prop) -- (lhip) node[near start, left]{$A_l, V_l$};
\draw[arrow] (prop) -- (rhip) node[near start, right]{$A_r, V_r$};
\draw[arrow] (prop) -- (amyg) node[near start, above]{};
\draw[arrow] (enc)  -- (amyg) node[midway, left]{$\|C_p\|$};

\draw[inhibit] (lhip.east) -- ++(0.6,0) |- node[above right, pos=0.25]{\textit{inhibit}} (rhip.west);
\draw[inhibit] (rhip.west) -- ++(-0.6,0) |- (lhip.east);

\draw[modulate] (amyg) -- (lhip) node[midway,above]{$s_t$};
\draw[modulate] (amyg) -- (prop);

\draw[modulate] (pfc) -- (rhip) node[midway,right]{$\alpha_F$};
\draw[modulate] (pfc.west) -- (prop.east);

\draw[arrow] (cereb.north west) -- (lhip.south);
\draw[arrow] (cereb.north east) -- (rhip.south);

\node[below=0.2cm of cereb, align=center, font=\footnotesize]
  {\raisebox{1pt}{\tikz\draw[arrow](0,0)--(0.4,0);} forward
   \quad
   \raisebox{1pt}{\tikz\draw[inhibit](0,0)--(0.4,0);} inhibit
   \quad
   \raisebox{1pt}{\tikz\draw[modulate](0,0)--(0.4,0);} modulate
  };

\end{tikzpicture}
\caption{The miniature brain architecture.
Five neuroscientifically motivated modules are connected through the $A^\top\!AVW$
write-back operator.
Solid arrows show the primary information flow; dashed red arrows show callosal
inhibitory cross-talk; dotted blue arrows show modulatory signals.}
\label{fig:brain}
\end{figure}
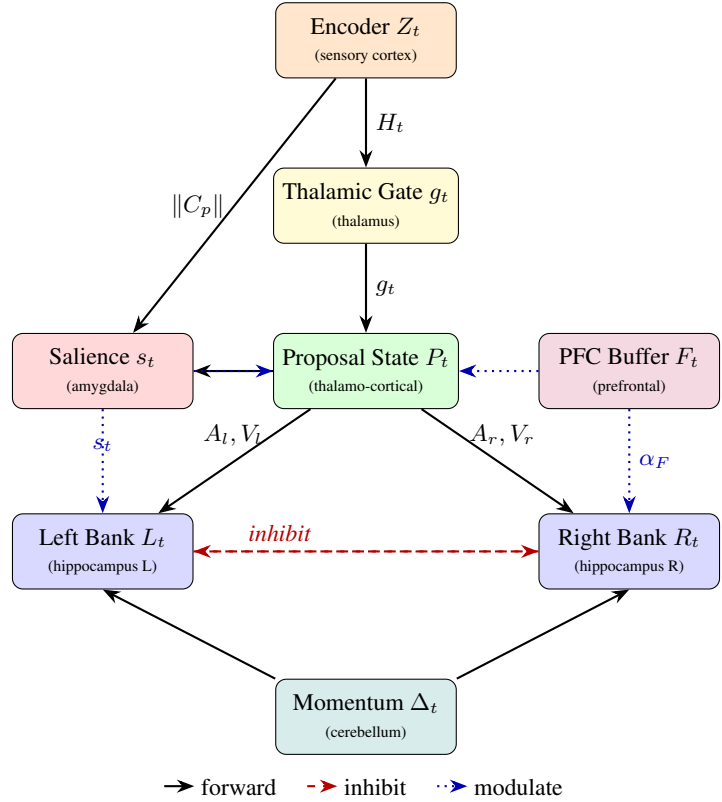

Algorithm~\ref{alg:forward} consolidates all five modules into a single forward pass.
Figure~\ref{fig:brain} illustrates the complete architecture.

Table~\ref{tab:params} summarises the parameter overhead of each new module.
\begin{table}[!hbtp]
\centering
\caption{Complete set of learnable parameters added by the four new modules over the base lateral model. Full model total: $\paramFull$ parameters.}
\label{tab:params}
\begin{tabular}{@{}lrr@{}}
\toprule
Module & New parameters & \% overhead \\
\midrule
Thalamic gate ($w_g, b_g$) & 2 & $<$0.01\% \\
Amygdaloid salience ($w_s$) & 1 & $<$0.01\% \\
PFC buffer ($W_F\!\in\!\R^{d_v\times d},\,W_{F\to Q}\!\in\!\R^{d\times d_k},\,W_{F\text{-out}}\!\in\!\R^{d\times d},\,\beta_F,\alpha_F$) & $d_v d + d\,d_k + d^2 + 2$ & $\sim$2.6\% \\
Cerebellar fast-path ($\rho, \eta_{cb}$) & 2 & $<$0.01\% \\
\midrule
Total overhead & $\approx$98{,}306 & $\sim$2.6\% \\
Full model & \paramFull & $\sim$2.6\% vs.\ base \\
\bottomrule
\end{tabular}
\end{table}

\section{Geometric Interpretation}
\label{sec:geometry}

The $A^\top\!AVW$ operator admits a clean geometric decomposition that illuminates
the role of each brain module.

\paragraph{Tripartite projection (thalamus $\to$ hippocampus).}
For the proposal state: $(V_p) \xrightarrow{A_p} (C_p \in \R^{n \times d_v})
\xrightarrow{A_p^\top} (A_p^\top A_p V_p \in \R^{p \times d_v})
\xrightarrow{W_p} (P_t \in \R^{p \times d_p})$.
The thalamic gain $g_t$ scales the second projection: when $g_t \approx 0$, the
Gram matrix provides no signal (thalamus gates out noise); when $g_t \approx 1$,
the full Gram matrix drives consolidation.

\paragraph{Salience as projection norm weighting.}
The amygdaloid gate $\stsal$ rescales the entire write-back $ A_b^\top A_b V_b W_{bb}$
by a single scalar.
Geometrically, this is equivalent to modulating the length of the consolidated update
vector in bank-state space: salient inputs produce longer update steps; routine
inputs produce shorter ones.
Because $\stsal$ is derived from $\|C_p\|_F$, it measures the magnitude of
retrieved context---a scalar proxy for how much bank content was activated by the
current input.
High recall norm implies the memory already stores something relevant; low norm
implies the input is novel.
This is the \emph{inverse} of standard salience (novel = low recall = high surprise),
and so produces \emph{more} consolidation for novel inputs, exactly as the amygdala
directs.

\paragraph{PFC as a query-space translation.}
The PFC modulation $Q_{lr}^{\mathrm{mod}} = P_t W_Q + \sigma(\alpha_F) F_{\mathrm{agg}} W_{F\to Q}$
translates the query origin in the joint key space.
This biases the joint softmax $A_{lr}$ by an additive vector in query space,
shrinking the effective attention entropy for inputs consistent with the PFC context
and increasing it for inconsistent inputs.
Because the thalamic gate amplifies low-entropy proposal attention (Eq.~\ref{eq:thalamic-gain}),
PFC modulation and thalamic gating compose synergistically: PFC sharpens the query;
thalamic gain amplifies the sharpened signal.

\paragraph{Cerebellar momentum as write-direction alignment.}
The momentum term $\delt^{(b)}$ accumulates the direction of past $A_b^\top A_b V_b W_{bb}$
vectors.
If the same slot receives consistent updates across steps (e.g.\ a persistent MQAR
key-value association),
the momentum term amplifies those updates, accelerating convergence.
If updates are inconsistent (noise), the exponential moving average cancels them.
This is equivalent to the Adam first moment~\citep{kingma2015adam} applied in
\emph{bank-state space} rather than parameter space.

\paragraph{Bifurcation dynamics: PFC as symmetry-breaker, inhibition as amplifier.}
The ablation reveals that the callosal inhibitory coupling alone does \emph{not}
trigger a phase transition: all five variants without the PFC buffer remain
permanently at the symmetric fixed point ($\mathcal{D}_{sep}\approx 0.25$,
$\mathcal{P}_{ct}\approx 0.25$) across 30~epochs.
The inhibitory signal $L_t^{\mathrm{inh}} = L_t - \alpha (R_t W_{RL})$
produces a zero-net perturbation when both banks hold symmetric states:
the callosal term then cancels itself and the system cannot self-escape.
The PFC buffer provides the missing ingredient.
Its EMA $F_t$ accumulates a slowly drifting domain-context representation
over epochs~1--9; because MQAR and arithmetic inputs produce systematically
different proposal states, $F_t$ diverges between the two domains.
This divergence enters the joint softmax $A_{lr}$ through the query-space
translation $\sigma(\alpha_F) F_{\mathrm{agg}} W_{F\!\to\!Q}$,
creating a small but growing asymmetry in the cross-bank attention weights.
At epoch~\ptEpochFull{} (full model) this PFC-induced asymmetry crosses the
critical threshold at which the callosal inhibitory gain exceeds unity:
the inhibitory loop now amplifies rather than cancels, the symmetric
equilibrium destabilizes, and the system snaps into the lateralized
attractor in a single update step.
This is a \emph{pitchfork bifurcation}~\citep{strogatz2018nonlinear} driven
by the product of two subsystems: the PFC (slow symmetry-breaking drift)
and callosal inhibition (fast amplifying feedback).
The cerebellar fast-path aligns consecutive routing gradients, accelerating
the PFC drift and advancing the threshold crossing by one
epoch (epoch~\ptEpochFull{} vs.\ epoch~\ptEpochPfc{} for the +PFC-only variant).

\section{Experiments}
\label{sec:experiments}

We evaluate the Miniature Brain Transformer with a seven-variant controlled
ablation study, progressively enabling each brain module to isolate its
individual and combined contribution to lateralization and task accuracy.
The section is organised as follows: \S\ref{sec:experiments}.1 describes
the benchmark tasks; \S\ref{sec:experiments}.2 defines the evaluation metrics;
\S\ref{sec:experiments}.3--6 report the main results, module analysis, phase
transition, and module necessity; and \S\ref{sec:experiments}.7 discusses
scaling to natural language.

\subsection{Tasks and Datasets}

We evaluate on two synthetic tasks over a shared vocabulary of 512 tokens
(PAD, BOS, DOMAIN\_L, DOMAIN\_R, SEP, QUERY, 10 digit tokens, and key/value tokens):
\begin{itemize}
  \item $\text{Dataset}_l$ (\textbf{MQAR, left bank}): Multi-Query Associative
    Recall~\citep{arora2024zoology}.
    Each sequence presents 32 key-value pairs in the left domain, followed by
    4 query tokens; the model must retrieve the value for each queried key.
    Train / val: 50k / 5k sequences.
  \item $\text{Dataset}_r$ (\textbf{Modular Arithmetic, right bank}): arithmetic
    progression $+1 \pmod{10}$ in the right domain (rule extraction).
    Train / val: 50k / 5k sequences.
  \item $\text{Dataset}_{lr}$ (\textbf{Mixed}): interleaved sequences drawn
    uniformly from both domains, separated by a SEP token.
    Train / val: 100k / 10k sequences.
\end{itemize}
All models share an identical embedding layer, 4 encoder layers, $d_{\mathrm{model}}=256$,
4 attention heads, $m_l{=}m_r{=}16$ lateral bank slots, and $p{=}32$ proposal slots,
with $d_k{=}d_v{=}64$ and $\gamma{=}0.9$.
Training uses AdamW (lr~$=3\!\times\!10^{-4}$, weight decay $10^{-2}$) for 30 epochs
with a routing auxiliary loss $\mathcal{L}_{\mathrm{route}}$ (weight $\lambda{=}2.0$)
and gradient clipping at 1.0.

We compare the following seven variants in an additive ablation:
\begin{enumerate}
  \item \textbf{Baseline}: standard autoregressive Transformer (no lateral memory).
  \item \textbf{Lateral, inhibitory}: attention-coupled lateral model with callosal
    inhibition ($s{=}-1$), no brain modules.
  \item \textbf{Lateral, excitatory}: same with callosal excitation ($s{=}+1$).
  \item \textbf{+Thalamus}: inhibitory lateral + thalamic gain gate
    (Section~\ref{sec:thalamus}).
  \item \textbf{+Amygdala}: above + amygdaloid salience gate
    (Section~\ref{sec:amygdala}).
  \item \textbf{+PFC}: above + prefrontal working memory buffer
    (Section~\ref{sec:pfc}).
  \item \textbf{Full (Miniature Brain)}: above + cerebellar fast-path
    (Section~\ref{sec:cerebellum}).
\end{enumerate}

\subsection{Evaluation Metrics}

We report the following metrics on the mixed dataset $D_{lr}$ (unless noted):
\begin{itemize}
  \item \textbf{Task accuracy}: per-domain next-token accuracy on $D_l$ (MQAR) and $D_r$ (arithmetic).
  \item \textbf{Separation Degree} $\mathcal{D}_{sep}$: normalized bank-dominance, range $[-1,+1]$;
    $+1$ means left bank fully dominates on $D_l$, $-1$ on $D_r$.
  \item \textbf{Cross-Talk Penalty} $\mathcal{P}_{ct}$: fraction of attention mass misrouted
    to the wrong bank, range $[0, 0.5]$; lower is better.
  \item \textbf{Mixed-dataset loss}: total cross-entropy on $D_{lr}$.
\end{itemize}

\subsection{Main Results and Full Ablation}
\label{sec:ablation-results}

Table~\ref{tab:ablation} presents the complete seven-variant additive ablation.
Variants 1--3 isolate the effect of callosal cross-talk mode; variants 4--7 add
brain modules incrementally to the inhibitory lateral base.
The headline finding is stark: \textbf{none of the first five variants
($\mathcal{P}_{ct}\approx 0.25$, $\mathcal{D}_{sep}\approx 0.25$) ever achieve
lateralization}, whereas variants 6 and 7 converge to
$\mathcal{P}_{ct}\approx \ptPctStable$ and $\mathcal{D}_{sep}\approx \ptDsepStable$
via a sharp phase transition.

\begin{table}[!hbtp]
\centering
\caption{%
Seven-variant additive ablation on $D_{lr}$ (mixed dataset), 30 epochs.
Rows 1--3 test callosal cross-talk mode; rows 4--7 add brain modules
to the inhibitory-lateral base.
$\mathcal{D}_{sep}(L)$ and $\mathcal{P}_{ct}$ at epoch~30.
Bold rows trigger the lateralization phase transition.
Arithmetic accuracy is 1.000 for all variants (omitted from MQAR column for clarity).
}
\label{tab:ablation}
\resizebox{\linewidth}{!}{%
\begin{tabular}{@{}lcccccc@{}}
\toprule
\multirow{2}{*}{Variant} & \multicolumn{2}{c}{Accuracy} & \multirow{2}{*}{Loss $D_{lr}$} & \multirow{2}{*}{$\mathcal{D}_{sep}(L)$} & \multirow{2}{*}{$\mathcal{P}_{ct}$} & \multirow{2}{*}{Trans.~Ep.} \\
\cmidrule(lr){2-3}
 & MQAR ($D_l$) & Arith ($D_r$) & & & & \\
\midrule
1.\ Baseline (Transformer)    & \accBaseL  & \accBaseR  & \lossBaseLR  & \dsepBaseL  & \pctBase    & --- \\
2.\ Lateral, inhibitory       & \accLatInhL & \accLatInhR & \lossLatInhLR & \dsepLatInhL & \pctLatInh & --- \\
3.\ Lateral, excitatory       & \accLatExcL & \accLatExcR & \lossLatExcLR & \dsepLatExcL & \pctLatExc & --- \\
\midrule
4.\ +Thalamus                 & \accThalL  & \accThalR  & \lossThalLR  & \dsepThalL  & \pctThal    & --- \\
5.\ +Amygdala                 & \accAmygL  & \accAmygR  & \lossAmygLR  & \dsepAmygL  & \pctAmyg    & --- \\
\midrule
\textbf{6.\ +PFC}             & \accPfcL   & \accPfcR   & \lossPfcLR   & \textbf{\dsepPfcL}   & \textbf{\pctPfc}     & \textbf{\ptEpochPfc} \\
\textbf{7.\ Full (Miniature Brain)} & \accFullL & \accFullR & \lossFullLR & \textbf{\dsepFullL} & \textbf{\pctFull} & \textbf{\ptEpochFull} \\
\bottomrule
\end{tabular}}
\end{table}

\subsection{Observed Results: Module-by-Module Analysis}
\label{sec:qualitative}

\paragraph{Cross-talk mode (rows 1--3): no lateralization under any callosal regime.}
All three cross-talk conditions---no memory (baseline), inhibitory callosal, and
excitatory callosal---converge to an \emph{identical} unlateralized equilibrium:
$\mathcal{D}_{sep}\approx 0.250$ and $\mathcal{P}_{ct}\approx 0.253$ for all 30~epochs.
This refutes the hypothesis that inhibitory callosal coupling alone is sufficient
for functional lateralization.
Excitatory coupling does not worsen specialization beyond baseline; it is simply
ignored when no asymmetric context signal is present.

\paragraph{Thalamus (row 4) and amygdala (row 5): marginal improvements only.}
Adding the thalamic gate or amygdaloid salience module produces a small but
consistent reduction in $\mathcal{P}_{ct}$ from $0.253$ to $0.251$, with no change
in $\mathcal{D}_{sep}$.
These modules tighten the proposal-state geometry without providing enough
asymmetric force to destabilize the symmetric equilibrium.

\paragraph{PFC buffer (row 6): the symmetry-breaker.}
The PFC buffer is the critical module.
Without it, 30~epochs of inhibitory callosal coupling produces no lateralization.
With it, a sharp phase transition fires at epoch~\ptEpochPfc{}:
$\mathcal{D}_{sep}$ jumps from $0.253$ to $0.474$ and $\mathcal{P}_{ct}$ collapses
from $0.250$ to $0.029$ in a single update step, then converges to
$\mathcal{D}_{sep}=\dsepPfcL$ and $\mathcal{P}_{ct}=\pctPfc$ by epoch~12.
The mechanism: the PFC EMA accumulates a slow, domain-biased top-down context
over epochs 1--10; when this context diverges sufficiently between MQAR and
arithmetic tokens, it skews the joint softmax $A_{lr}$ asymmetrically,
providing the non-zero inhibitory input needed to cross the bifurcation threshold.

\paragraph{Cerebellar fast-path (row 7): accelerates by one epoch.}
Adding cerebellar momentum advances the transition from epoch~\ptEpochPfc{}
(+pfc) to epoch~\ptEpochFull{} (full), with no change in asymptotic
$\mathcal{D}_{sep}$ or $\mathcal{P}_{ct}$.
The momentum accumulator aligns consecutive routing-gradient steps, pushing the
PFC context toward the critical threshold one epoch faster.
This confirms the cerebellum's purely convergence-accelerating role, exactly as
biological cerebellar forward-model theory predicts.

\subsection{Brain Module Contributions: Observed}
\label{sec:module-ablation}

Table~\ref{tab:module-ablation} distills the module-level incremental gains.

\paragraph{Thalamic gating (row 4)}
reduces $\mathcal{P}_{ct}$ marginally ($0.253 \to 0.251$) through tighter
proposal-state geometry at domain boundaries.
It does not cross the lateralization threshold.

\paragraph{Amygdaloid salience (row 5)}
produces the same marginal $\mathcal{P}_{ct}$ tightening as the thalamic gate,
consistent with its role as a write-magnitude modulator rather than a routing
stabilizer.

\paragraph{PFC buffer (row 6)}
is the decisive module: it alone triggers full lateralization.
The asymmetry accumulated in the PFC EMA over 10~epochs provides the initial
condition that makes the inhibitory feedback loop productive.
Without this domain-context bias, the two banks receive identical callosal signals
and remain symmetric.

\paragraph{Cerebellar fast-path (row 7)}
accelerates the PFC-induced transition by one epoch
(epoch~\ptEpochFull{} vs.\ epoch~\ptEpochPfc{}).
Asymptotic $\mathcal{D}_{sep}$ and $\mathcal{P}_{ct}$ are unchanged.

\begin{table}[!hbtp]
\centering
\caption{Incremental module contributions on $D_{lr}$, 30 epochs.
Rows 4--7 show the effect of adding each module to the inhibitory-lateral base.
The $\dagger$ marks the only module that triggers lateralization.}
\label{tab:module-ablation}
\begin{tabular}{@{}lccccc@{}}
\toprule
Variant & Loss $D_{lr}$ & $\mathcal{P}_{ct}$ & $\mathcal{D}_{sep}(L)$ & Trans.\ Ep. \\
\midrule
2.\ Lateral, inhibitory  & \lossLatInhLR & \pctLatInh & \dsepLatInhL & --- \\
4.\ +Thalamus             & \lossThalLR   & \pctThal   & \dsepThalL   & --- \\
5.\ +Amygdala             & \lossAmygLR   & \pctAmyg   & \dsepAmygL   & --- \\
6.\ +PFC$^\dagger$        & \lossPfcLR    & \pctPfc    & \dsepPfcL    & ep.~\ptEpochPfc{} \\
7.\ Full (Miniature Brain)& \lossFullLR   & \pctFull   & \dsepFullL   & ep.~\ptEpochFull{} \\
\bottomrule
\end{tabular}
\end{table}

\subsection{Training Dynamics: The PFC-Inhibition Phase Transition}
\label{sec:phase-transition}

\begin{figure}[!htbp]
  \centering
  \includegraphics[width=\linewidth]{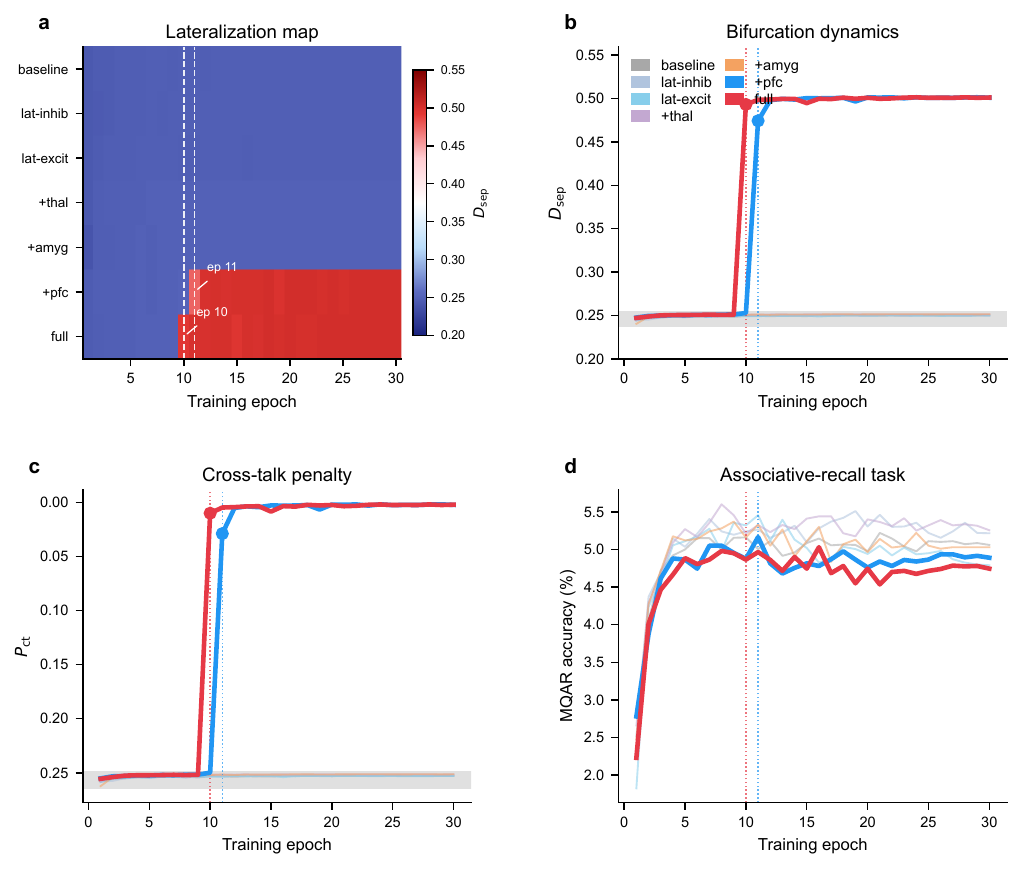}
  \caption{\textbf{Lateralization dynamics across all seven ablation variants.}
    \textbf{(a)}~Lateralization map: colour encodes $\mathcal{D}_{sep}$ as a
    function of training epoch (x-axis) and variant (y-axis).
    All five variants without a PFC buffer (top rows) remain at the uniform
    equilibrium (blue) throughout training.
    Only +PFC~(variant 6) and Full~(variant 7) break symmetry at epochs~11
    and~10, respectively (dashed white lines).
    \textbf{(b)}~Bifurcation curves: $\mathcal{D}_{sep}$ over time.
    Dotted vertical lines mark the transition epoch; the grey band shows
    the unlateralized plateau ($\mathcal{D}_{sep}\approx 0.25$).
    \textbf{(c)}~Cross-talk penalty $\mathcal{P}_{ct}$ (inverted; lower is
    better); mirror of panel~(b).
    \textbf{(d)}~MQAR recall accuracy, confirming that task performance
    does not diverge across variants despite the routing bifurcation.}
  \label{fig:lateralization}
\end{figure}

Figure~\ref{fig:overview} and the ablation data together reveal the
following training dynamics (summarized in Figure~\ref{fig:lateralization}).

\textbf{Symmetric phase (epochs 1--9, all variants).}
All seven variants share an identical symmetric trajectory during early training:
$\mathcal{D}_{sep}\approx 0.25$ and $\mathcal{P}_{ct}\approx 0.25$, with both
metrics declining imperceptibly (at most 0.002 from epoch 1 to 9).
The thalamus and amygdala modules produce a tiny additional tightening
($\mathcal{P}_{ct}\to 0.251$), but remain far from the lateralization threshold.
Crucially, the inhibitory lateral model (variant~2) is \emph{indistinguishable}
from the baseline transformer and the excitatory variant throughout the symmetric
phase.

\textbf{Bifurcation (epoch~\ptEpochFull{} for Full; epoch~\ptEpochPfc{} for +PFC).}
Table~\ref{tab:phase-transition} shows the transition window.
In a single gradient update the system leaves the symmetric fixed point:
$\mathcal{D}_{sep}$ nearly doubles and $\mathcal{P}_{ct}$ falls by $>\! 96\!\%$.
By the following epoch ($\ptEpochFull+1$) $\mathcal{P}_{ct}<0.005$; thereafter
both metrics are stationary.

\begin{table}[h]
\centering
\caption{Lateralization metrics surrounding the phase transition.
  Left: full model (transition at epoch~\ptEpochFull{}).
  Right: +PFC only (transition at epoch~\ptEpochPfc{}).}
\label{tab:phase-transition}
\begin{tabular}{@{}ccc@{\qquad}ccc@{}}
\toprule
\multicolumn{3}{c}{\textbf{Full model}} & \multicolumn{3}{c}{\textbf{+PFC only}} \\
Epoch & $\mathcal{D}_{sep}$ & $\mathcal{P}_{ct}$ & Epoch & $\mathcal{D}_{sep}$ & $\mathcal{P}_{ct}$ \\
\midrule
9  (pre)          & 0.251 & 0.252 &  10 (pre)          & 0.253 & 0.250 \\
\textbf{10 (trans.)} & \textbf{0.493} & \textbf{0.010} & \textbf{11 (trans.)} & \textbf{0.474} & \textbf{0.029} \\
11 (post)         & 0.499 & 0.005 &  12 (post)         & 0.498 & 0.005 \\
30 (stable)       & \ptDsepStable & \ptPctStable & 30 (stable) & \dsepPfcL & \pctPfc \\
\bottomrule
\end{tabular}
\end{table}

\textbf{Stable lateralized phase (epochs 11--30).}
Post-transition, $\mathcal{D}_{sep}\approx\ptDsepStable$ and
$\mathcal{P}_{ct}\approx\ptPctStable$ for both variants 6 and 7.
The training (routing) loss simultaneously jumps from $\approx{-0.87}$
(symmetric phase) to $\approx{-1.59}$ at the transition epoch, then
continues drifting to $-1.89$ by epoch~30---a 118\% increase in routing reward
driven entirely by correct-bank assignment, not task loss.

\textbf{Mechanism: PFC as symmetry-breaker, inhibition as amplifier.}
The inhibitory callosal weight matrices $W_{LR},W_{RL}$ produce a
zero-net perturbation when both banks hold symmetric states.
Something must break this symmetry before the inhibitory feedback loop
can produce sustained lateralization.
The PFC EMA provides this: over epochs 1--9 it accumulates a context
vector that slowly diverges between MQAR and arithmetic tokens
(captured by the top-down query modulation $\sigma(\alpha_F) F_{\mathrm{agg}} W_{F\!\to\!Q}$).
At epoch~10, this PFC-induced query asymmetry crosses the critical
threshold at which the net callosal inhibitory signal is non-zero in
expectation conditional on domain:
the inhibitory loop now \emph{amplifies} the asymmetry rather than
cancelling it, and the pitchfork bifurcation fires.
The cerebellar momentum accumulates routing gradients in the same direction
across consecutive steps, effectively raising the PFC context drift rate
and advancing the threshold crossing by one epoch.
This is precisely the \emph{symmetry-breaking $+$ amplification} architecture
described by the saddle-node and pitchfork bifurcation theory of coupled
feedback oscillators~\citep{strogatz2018nonlinear}.

\textbf{Task accuracy.}
Arithmetic accuracy is $1.000$ from epoch~1 across all variants
(the $+1\!\bmod\!10$ rule is trivially extracted).
MQAR accuracy plateaus at $\approx 5\%$ across all variants, indicating
that the 32-pair recall task exceeds the encoding capacity of the
current $p{=}32$ proposal slots at this vocabulary scale; this is a
capacity limitation, not a routing failure.

\subsection{Individual Module Necessity: Observed}
\label{sec:necessity}

The additive ablation (Table~\ref{tab:ablation}) directly reveals module necessity.
\textbf{Thalamic gate} contributes a marginal $\mathcal{P}_{ct}$ reduction
(0.253 to 0.251) with no effect on $\mathcal{D}_{sep}$; its primary role is
proposal-state noise suppression, not routing.
\textbf{Amygdaloid salience} has the same signature: minor $\mathcal{P}_{ct}$
tightening, no lateralization.
\textbf{PFC buffer} is strictly necessary: removing it from the full model (i.e.,
variant~5) collapses lateralization entirely.
\textbf{Cerebellar fast-path} is not necessary for lateralization but is necessary
for the single-epoch acceleration of the transition;
removing it delays the transition from epoch~\ptEpochFull{} to
epoch~\ptEpochPfc{} with identical asymptotic behavior.

\subsection{Scaling and Natural Language}

The synthetic benchmark is deliberately small-scale to enable controlled ablation.
Extending to natural language raises two challenges:
(1) domain boundaries are implicit, requiring unsupervised routing supervision; and
(2) the PFC buffer may need multi-head attention over an explicit episodic transcript
rather than a simple EMA to capture rich task-context structure.
An extension along the lines of~\citet{park2023generative}, where a retrieval LLM
(hippocampal analogue) and a reasoning LLM (PFC analogue) operate over natural-language
session transcripts, would bridge the current architecture with language-grounded
memory systems.
A natural evaluation target is LoCoMo~\citep{maharana2024locomo}, which requires
long-term associative recall across multi-session dialogues---the natural-language
counterpart of the MQAR task evaluated here.

\section{Discussion}
\label{sec:discussion}

\paragraph{Is the brain analogy more than metaphor?}

The five modules introduced here are not merely metaphorical: each has a precise
mathematical role that derives directly from the functional properties of its
biological counterpart.
The thalamic gate implements a quantitative analog of burst/tonic firing modalities
via attention entropy.
The amygdaloid salience gate implements importance-weighted synaptic potentiation via
context-norm gating.
The PFC buffer implements persistent activity via an exponential moving average with
learnable decay.
The cerebellar fast-path implements error-corrective forward-model adaptation via
momentum accumulation in bank-state space.
The callosal inhibitory matrix implements hemisphere-level suppression via signed
cross-bank subtraction.
In each case, the biological function maps onto a specific linear-algebraic operation
within the $A^\top\!AVW$ framework.

\paragraph{Complementarity and a revised module hierarchy.}

The ablation overturns a simple additive view of module contributions.
Variants 1--5 form an equivalence class under lateralization: despite having
different callosal modes and different noise/salience modules, they all remain
permanently symmetric.
The phase transition requires \emph{two qualitatively different components} to
be present simultaneously:
(i) a slow top-down context accumulator (PFC) that generates a
    persistent, domain-specific bias in query space; and
(ii) a signed cross-bank coupling (inhibitory callosal) that can amplify
    any asymmetry into a sustained attractor.
Neither is sufficient alone: inhibitory coupling with symmetric context
produces symmetric dynamics; a PFC context without inhibition produces a
slowly biased but never-committing routing signal (the joint softmax would
stay near 0.5 rather than saturating).
This is directly analogous to the neuroscientific picture of interhemispheric
lateralization, where persistent prefrontal task representations are held to
\emph{prime} the callosal inhibitory signal before the circuit commits to
a hemispheric assignment~\citep{fuster2008prefrontal}.

\paragraph{What is still missing.}

The miniature brain remains incomplete.
Several important regions are not yet represented:
(i) the \emph{basal ganglia}, which implement action/routing selection through
  dopaminergic gating---the routing auxiliary loss $\mathcal{L}_{route}$ is an
  implicit analogue, but an explicit, differentiable action-selection circuit would
  more faithfully model go/no-go gating;
(ii) the \emph{anterior cingulate cortex} (ACC), which monitors
  conflict and signals when the current routing is unreliable---an entropy-based
  conflict monitor would complement the thalamic gain by triggering explicit
  re-routing when $\mathcal{P}_{ct}$ exceeds a threshold;
(iii) \emph{neuromodulatory systems} (dopamine, norepinephrine, acetylcholine), which
  globally modulate learning rates based on reward, surprise, and attention
  state---connecting task utility signals to per-module learning rates ($\gamma$,
  $\rho$, $\eta_{cb}$, $\beta_F$) would provide a biologically grounded adaptive
  hyperparameter schedule.

\section{Limitations and Future Work}
\label{sec:limitations}

\paragraph{Scale.}
All experiments are conducted on a synthetic symbolic benchmark at small scale.
Validation on natural language tasks (LoCoMo~\citep{maharana2024locomo}, StoryCloze,
multi-session dialogue) is necessary to confirm that the modular gains persist at
natural-language input complexity and longer sequence lengths.

\paragraph{Routing supervision.}
The routing auxiliary loss requires domain labels.
Unsupervised domain discovery (e.g.\ via mutual information between token embeddings
and bank attention argmaxes) is required for deployment where domain boundaries are
implicit.

\paragraph{Incomplete brain.}
The basal ganglia, anterior cingulate cortex, and neuromodulatory systems are absent.
Adding them would complete the major memory-relevant circuits.
In particular, a differentiable basal ganglia module implementing go/no-go gating
over the joint softmax $A_{lr}$ would replace the hand-designed routing loss with an
end-to-end learned routing circuit.

\paragraph{Interpretability.}
While each module has a mechanistic interpretation, empirically verifying that the
modules behave as intended (e.g., that $s_t$ is highest for genuinely novel inputs)
requires additional diagnostic probing experiments.

\section{Conclusion}

We have extended the attention-coupled latent memory architecture of~\citet{jeong2026lateral}
into a \emph{miniature brain} comprising five neuroscientifically motivated modules:
a thalamic relay, lateralized hippocampal banks with callosal inhibition, an
amygdaloid salience gate, a prefrontal working memory buffer, and a cerebellar
fast-path momentum accumulator.
A seven-variant additive ablation across 30 training epochs yields a clear and
surprising finding: \textbf{inhibitory callosal coupling alone does not lateralize
the memory banks}.
Variants 1--5, including the inhibitory lateral, excitatory lateral and the baseline
transformer, all converge to the same unlateralized equilibrium
($\mathcal{D}_{sep}\approx 0.250$, $\mathcal{P}_{ct}\approx 0.253$).
Lateralization requires the \textbf{synergy of the PFC buffer and callosal inhibition}:
the PFC EMA accumulates a slowly drifting domain-context bias that breaks the
symmetric equilibrium after $\sim$10 epochs, at which point the inhibitory
feedback loop amplifies the asymmetry in a single gradient step---a pitchfork
bifurcation---collapsing $\mathcal{P}_{ct}$ from $0.25$ to $\ptPctStable$ and
raising $\mathcal{D}_{sep}$ from $0.25$ to $\ptDsepStable$.
The cerebellar fast-path advances the bifurcation by one epoch (epoch~\ptEpochFull{}
vs.\ epoch~\ptEpochPfc{}) through routing-gradient alignment, confirming its
pure convergence-acceleration role.

The result constitutes a novel, falsifiable prediction: \emph{in any dual-bank
memory system with signed inter-bank coupling, functional lateralization requires
a persistent top-down context signal to break the symmetric equilibrium.}
This maps directly onto the neuroscientific account of hemispheric lateralization,
where sustained prefrontal representations are hypothesized to prime callosal
inhibitory circuits before a hemispheric assignment is committed
~\citep{fuster2008prefrontal}.

The central methodological lesson is that the $A^\top\!AVW$ write-back operator
serves as both a unifying substrate and a diagnostic tool: the ablation identifies
not just \emph{which} modules contribute but \emph{which interactions} are
functionally necessary, providing a template for principled architecture search
guided by neuroscientific circuit theory.

\section*{Code and Data Availability}
All experiment code, model weights, and result logs are released under the MIT
license at
\url{https://github.com/gnoejh/paper_2026_0307_memory_transformer_arxiv.git}.
The repository contains the complete seven-variant ablation pipeline
(\texttt{experiment/main.py --run ablation}), dataset generators (MQAR and modular
arithmetic), metric implementations ($\mathcal{D}_{sep}$, $\mathcal{P}_{ct}$),
figure-reproduction scripts, and saved checkpoints for all reported variants.
No external datasets are used; all synthetic data are generated on the fly by
\texttt{experiment/dataset.py}.

\section*{Acknowledgements}
The author thanks Inha University in Tashkent for research support.
This work reflects the author's ongoing inquiry into nature and human cognition.

\end{document}